\documentstyle{article}

\def\be{\begin{equation}}
\def\ee{\end{equation}}
\def\bea{\begin{eqnarray}}
\def\eea{\end{eqnarray}}
\begin{document}

\pagestyle{empty}
\vskip-10pt
\vskip-10pt
\hfill {\tt hep-th/0110248}
\begin{center}
\vskip 3truecm
{\Large\bf
The $d=6$, $(2,0)$-tensor multiplet coupled to self-dual strings
}\\ 
\vskip 2truecm
{\large\bf
Andreas Gustavsson
}\\
\vskip 1truecm
{\it Institute of Theoretical  Physics,
Chalmers University of Technology, \\
S-412 96 G\"{o}teborg, Sweden}\\
\vskip 5truemm
{\tt f93angu@fy.chalmers.se}
\end{center}
\vskip 2truecm
\noindent{\bf Abstract:} 
We show that the central charges that group
theory allows in the $(2,0)$-supersymmetry translations algebra arise
from a string and a 3-brane by commuting two supercharges. We show
that the net force between two such parallel strings vanishes. We show
that all the coupling constants are fixed numbers, due to
supersymmetry, and self-duality of the three-form field strength. We
obtain a charge quantization for the self-dual field strength, and
show that when compactifying on a two-torus, it reduces to the usual
quantization condition of $N=4$ SYM with gauge group $SU(2)$, and with
coupling constant and theta angle given by the $\tau$-parameter of the
two-torus, provided that we pick that chiral theory which corresponds
to a theta function with zero characteristics, as expected on
manifolds of this form.

\vfill \vskip4pt

\eject
\newpage
\pagestyle{plain}

\section{Introduction}
 
It is believed that $N=4$ super Yang-Mills theories in four dimensions
has its origin in $(2,0)$ supersymmetric six-dimensional theories
\cite{Witten95} \cite{Verlinde}. The S-duality property of the $N=4$
theory would then have a purely geometrical explanation as being the
modular group of a two-torus when compactifying the six-dimensional
theory to four dimensions. It is however not possible to proceed
straightforwardly and reduce the action of the six-dimensional theory
since there does not exist any covariant action for a self-dual
three-form field strength which is in the $(2,0)$ tensor
multiplet. What one can do is to reduce the equations of motion. By
integrating the self-dual field strength (divided by some number) over
spatial three-cycles in a non-trivial topology one gets a quantity
which is either integer or integer shifted by $1/2$, depending on
which chiral theory one has. The various chiral theories can be
labeled by the characteristics $\alpha$ and $\beta$ in
$(\frac{1}{2}{\bf{Z}})^{\frac{1}{2}b_3}$, where $b_3$ is the dimension
of the third homology group of the six-manifold or the third Betti
number. One can reduce the self-dual gauge potential to four
dimensions. Due to its self-duality it reduces to one (compact) scalar
and one gauge field. By compactifying on a $T^2$ with modular
parameter $\tau$ one should get the charge quantization in four
dimensions with a theta-angle $\theta$ \cite{Witten79}. The Yang-Mills
coupling constant $g_{YM}$ and the theta angle should combine to the
$\tau$ of the $T^2$ as $\tau=\frac{\theta}{\pi}+\frac{8\pi
i}{g_{YM}^2\hbar}$. We will see that this is true, but only for the
the theory with zero characteristics. This is in agreement with the
observation that on manifolds with one circle being time and one (or
several, in this case two) one-cycle(s) being time-like, the only
theory which can candidate to give a modular invariant partition
function is that with zero characteristics \cite{G},
\cite{Bonelli}. One should perhaps not expect full modular invariance
of the partition function only for the tensor part, but one should
expect that this partition function transform to itself at least up to
a phase factor and that is the case only for the theory with zero
characteristics.

The free $(2,0)$-theory has no adjustable parameters. Their numerical
values are determined from the $(2,0)$-supersymmetry up to an overall
coupling constant. This overall coupling constant, which we will call
$\lambda$, can only take one particular value, but that does not
follow from supersymmetry. We have found two seemingly unrelated ways
to determine its value, or more precisely, the ratio $\lambda/g$ where
$g$ is the unit in which the self-dual charges are quantized. The
first criterion is that there should only be finitely many chiral
theories. The second criterion is that the Wilson surface observables,
$\exp{2\pi i\int_D \frac{H^+}{g}}$, over three-dimensional surfaces
$D$, should commute, in order for the $U(1)$ Wilson and 't Hooft lines
which one obtains when reducing to four dimensions, to commute. We do
not know how to write an observable in six dimensions that reduces to
$SU(N)$ Wilson and 't Hooft lines in four dimensions.

In section 2 we examine how $(2,0)$ supersymmetry constrains the
parameters in an action. We write an action for a non-self-dual gauge
field, from which the equation of motion for the self-dual part,
$H^+$, of the field strength can be obtained by decomposing $H$ as
$H=H^+ + H^-$. Supersymmetry fixes the sizes of the parameters in this
action only up to an overall factor, which, as we will see in section
3 and 4, is determined from the self-duality of the field strength. We
construct supercharges out of the fields in the $(2,0)$ tensor
multiplet. When we anti-commute two supercharges, in the same manner
as in \cite{OW}, we find central charges which correspond to a string
and a 3-brane, respectively. We use the BPS-condition on the string
tension to fix the relative size of the constants in the action which
describes a tensor multiplet that couples to strings. We show that the
net force between two equally charged parallel strings vanishes due to
attraction via scalars and repulsion due to the self-dual tensor field.

In section 3 we examine how self-duality of the field strength
constrains the value that the coupling constant takes, given a time
direction. The natural framework for this is the Hamiltonian
formulation. The condition we want to satisfy is that the partition
function for the non-chiral two-form potential shall be possible to
holomorphically factorize into $\frac{1}{2}{b_3}$ number of terms.

In section 4 we show that the value we have obtained of the coupling
is precisely that which gives the `correct' commutation relations of
the Wilson surface observables.

In section 5 we obtain the usual quantization conditions with a
theta-angle of $N=4$ SYM with gauge group $SU(2)$ spontaneously broken
to $U(1)$ by compactifying a $(2,0)$-theory with one massless tensor
multiplet and with zero characteristics, on $T^2\times M_4$, where
$M_4=S^1\times M_3$. The $S^1$ is time.

When we had finished this work we got informed that results of section
$4$ has also been obtained in \cite{GLPT}.

\section{Coupling of the tensor multiplet to a classical string}

In this section we will assume that we have a flat six-dimensional
background with metric $G_{\mu \nu}$ = diag$(-1,1,1,1,1,1)$. We will
use $\mu=\{0,i\}=0,1,...5$ as vector indices and $A,B,...$ as Dirac
spinor indices of the Dirac representation $8=4\oplus 4'$, and
$\alpha,\beta,...$ and $\alpha',\beta',...$ as the Weyl spinor indices
respectively, in the Lorentz group $SO(1,5)$; $a,b,...=1,2,...,5$ as
vector indices and $i,j,...$ as spinor indices in the R-symmetry group
$SO(5)_R$. More conventions about our spinors are found in the
appendix $A$. We define the three-form field strength $H$ from the
two-form gauge potential $B$ as $H=dB$. Here $H=\frac{1}{3!}H_{\mu \nu
\rho} dx^{\mu}\wedge dx^{\nu}\wedge dx^{\rho}$ and
$dB=\frac{1}{2!}\partial_{\mu}B_{\nu \rho}dx^{\mu}\wedge
dx^{\nu}\wedge dx^{\rho}$. The components of the field strength are
thus \be H_{\mu \nu \rho}=3\partial_{[\mu}B_{\nu
\rho]}=\partial_{\mu}B_{\nu \rho} + \partial_{\rho}B_{\mu \nu} +
\partial_{\nu}B_{\rho \mu} \ee We note that there does not exist a
decomposition of the gauge potential $B$ into chiral potentials
$B^{\pm}$ unless the fields satisfy the equation of motion. If $H=dB^+
+ dB^-$ where $*dB^{\pm}=\pm dB^{\pm}$ then $d*H=dH=0$. Conversely, if
$dH=d*H=0$, then we can locally write $H=dB$ and $*H=d\tilde{B}$ and
hence, locally, we have that $H=dB^+ + dB^-$ where we can take
$B^{\pm}=\frac{1}{2}(B\pm \tilde{B})$. We will in this paper always
assume that the fields are on-shell so that such chiral potentials
exist (locally).

The supersymmetry charges of the $d=6$, $(2,0)$-theory transform in
the representation $(4,4)$ of $SO(1,5)\times SO(5)$. The
anti-commutator of two such supercharges will transform in the
representation (s (a) means the (anti) symmetric part) \bea
&&((4,4)\times(4,4))_{s}\simeq (6_a\oplus 10^+_s,1_a\oplus 5_a\oplus
10_s)_s\cr &&=(6_a,1_a)\oplus (6_a,5_a)\oplus (10^+_s,10_s)\simeq
P_{\mu}\oplus Z_{\mu a}\oplus W^+_{\mu\nu\rho, a b}, \eea so the most
general $SO(1,5)\times SO(5)_R$-invariant supertranslations algebra is
\cite{HLW} 
\bea \{Q_{\alpha i},Q_{\beta j}\} & = &
i\Big(\Omega_{ij}(\gamma^{\mu})_{\alpha\beta}P_{\mu}+(\sigma^a)_{ij}
(\gamma^{\mu})_{\alpha\beta}Z_{\mu
a}\cr
&&+\frac{1}{2!3!}(\sigma^{ab})_{ij}(\gamma^{\mu\nu\rho})_{\alpha\beta}W^+_{\mu\nu\rho,ab}\Big).
\eea 
The overall factor $i$ in the right hand side comes from the
symplectic Majorana condition $(Q_{\beta
j})^{\dag}=i\Omega^{ji}Q_{\alpha i}(\gamma^0)_{\alpha\beta}$. We
define the translation generator as
$[P_{\mu},\cdot]=i\partial_{\mu}$. From this algebra one derives that
there is a massless tensor multiplet on which these supercharges act
as \cite{CKP} \bea [Q_{\alpha i},B_{\mu\nu}] &=&
i(\gamma_{\mu\nu})_{\alpha}{}^{\beta}\psi_{\beta i}\cr [Q_{\alpha
i},\phi_a] &=& i(\sigma_a)_{i}{}^j\psi_{\alpha j}\cr \{Q_{\alpha
i},\psi_{\beta j}\} &=& {{i}\over
{24}}\Omega_{ij}(\gamma^{\mu\nu\rho})_{\alpha\beta}H^+_{\mu\nu\rho}+{{i}\over
{2}}(\sigma^a)_{ij}(\gamma^{\mu})_{\alpha\beta}\partial_{\mu}\phi_a.\label{susy}
\eea The commutator of two variations close only if one uses the
equations of motion. The action for this massless multiplet can be
determined by requiring that the supercharges transform the massless
fields in the tensor multiplet as above. We find that the supercharges
\cite{BSP} 
\be 
Q_{\alpha i}
= {1\over 6}\int d^5x\gamma^{\mu\nu\rho}\gamma^0\psi_{\alpha i}H^+_{\mu\nu\rho} 
+ 2\int d^5x\sigma_a\gamma_{\mu}\gamma^0\psi_{\alpha i}\partial^{\mu}\phi^a 
\ee
will do the job if and only if the canonical equal-time commutation
relations are 
\bea 
[\phi^a(x),\partial^0 \phi_b(y)] &=&
-i\frac{1}{2}\delta_b^a\delta^5(x-y)\cr \{\psi_{\alpha
i}(x),\psi_{\beta j}(y)\} &=&
i\frac{1}{4}\Omega_{ij}(\gamma_0)_{\alpha\beta}\delta^5(x-y)\cr
[H_{lmn}^+(x),H_{ijk}^+(y)] &=&
i{3\over{2}}\epsilon_{[ijklm}\partial_{n]}\delta^5(x-y) 
\eea 
which in turn can be derived from the non-chiral action 
\be 
S = \int
d^6x(-\frac{1}{12}H_{\mu\nu\rho}H^{\mu\nu\rho} -
\partial_{\mu}\phi^a\\\partial^{\mu}\phi_a + 4 \psi^{\alpha
i}\Omega_{ij}(\gamma^{\mu})_{\alpha\beta}\partial_{\mu}\psi^{\beta
j}).\label{action} 
\ee 
(The last of these commutation relations is a
bit tricky and is derived in appendix $B$.) We will now anti-commute
two such supercharges and pay attention only to terms that survive
only on topologically non-trivial six-manifolds \cite{OW}, which will
turn out to correspond to the non-compact topologies one gets by
deleting an infinite string and a 3-brane respectively from the
M5-brane world-volume which we have assumed to be flat, i.e. with
vanishing intrinsic curvature \cite{GGT}. (The extrinsic curvature,
that is, how the M5-brane is embedded in eleven dimensions, is an
other thing which we don't consider here.) We notice that
$\gamma^{\mu\nu\rho}H^+_{\mu\nu\rho}=2\gamma^{ijk}H^+_{ijk}$ due to
self-duality. Then we get 
\bea 
\{Q_{\alpha i},Q_{\beta j}\}&=&...+{i\over 3}(\sigma^a)_{ij}
(\gamma^{ijkl0})_{\alpha\beta}\int
d^5x H_{ijk}\partial_{l}\phi_a\cr
&&+i(\sigma^{ab})_{ij}(\gamma^{\mu\nu 0})_{\alpha\beta}\int
d^5x \partial_{\mu}\phi_a\partial_{\nu}\phi_b\cr
&=&...+i(\sigma^a)_{ij}(\gamma_{m})_{\alpha\beta} 2 \int  H\wedge
d\phi_a\wedge dx^m\cr
&&+\frac{i}{2!3!}(\sigma^{ab})_{ij}(\gamma_{klm})_{\alpha\beta} 4 \int
d\phi_a\wedge d\phi_b\wedge dx^k\wedge dx^l\wedge dx^m 
\eea 
Now we assume that we have an infinite string $\Sigma$ in the
$X^5$-direction, located at $X^1=X^2=X^3=X^4=0$, so that we integrate
over the manifold
${\bf{R}}^5-{\bf{R}}={\bf{R\times(R}}^4-\{0\})={\bf{R\times(R_+\times}}S^3)$. Then we have 
\bea
&&=...+i(\sigma^a)_{ij}(\gamma_5)_{\alpha\beta}\int_{string}dx^5 2 \int_{S^3\times{\bf{R_+}}}H^+\wedge d\phi_a\cr
&&=...+i(\sigma^a)_{ij}(\gamma_5)_{\alpha\beta} 2 \int_{string}dx^5g^+\phi_a\left.\right|_{\Sigma}
\eea 
where in the last step we have defined $g^+\equiv \int_{S^3}H^+$. 
From this we read off  
\be 
Z_{\mu a}=2 \int_{string}dx^5g^+\delta_{\mu}^5\phi_a\left.\right|_{\Sigma}.
\ee 
The string tension $T$ is given by  
\be
\int_{string}T=\sqrt{Z_{\mu a}Z^{\mu a}} 
\ee 
for BPS-saturated
strings, so 
\be 
T = 2 g^+ \sqrt{\phi^a\phi_a}\left.\right|_{\Sigma}.  
\ee
The tension will contain a part coming from the $\phi$-field the string
produces itself, plus a part coming from $\phi$-fields produced by other 
strings.

Similarly for an infinite 3-brane in the $X^{1,2,3}$ directions,
localized at $X^4=X^5=0$, we get the manifold ${\bf{R^5-R^3=R^3\times
(R_+\times }} S^1)$, and 
\bea
=...+i\frac{1}{2!}(\sigma^{ab})_{ij}(\gamma_{123})_{\alpha\beta} 4 \int_{3-brane}
dx^1\wedge dx^2\wedge dx^3 \int_{S^1\times{\bf{R_+}}} d\phi_a\wedge
d\phi_b 
\eea 
from which we read off 
\be
W^{\mu\nu\rho}{}_{ab}=4 \delta^{\mu\nu\rho}_{123}\int_{3-brane}
dx^1\wedge dx^2\wedge dx^3 \int_{S^1\times{\bf{R_+}}} d\phi_a\wedge
d\phi_b 
\ee 
and the three-brane tension, 
\be
\tau_3= 4 \int_{S^1\times{\bf{R_+}}} d\phi_a\wedge d\phi_b.  
\ee

The equations of motion for the bosonic fields in the tensor multiplet
coupled to a string can be obtained by adding to the action the
interaction terms 
\be 
- \sum_{i} \int_{\Sigma_i}
T\sqrt{-h}d^2\sigma,\label{interaction} 
\ee 
where $i$ runs over all the string world-sheets, and $h_{\alpha\beta}$ 
is the induced metric
on the string world-sheet. For an infinite self-dual BPS-string
$\Sigma$ in the $X^5$-direction we get 
\be 
- \int_{\Sigma} T\sqrt{-h}d^2\sigma = -2g^+ \int d^6x
|\phi(x)|\int_{\Sigma}d^2\sigma\delta^6(x-X(\sigma)) 
\ee 
where $|\phi|\equiv \sqrt{\phi^a\phi_a}$ and if we e.g. assume that
$\phi_a=\delta_{a5}|\phi|$ then we get the equation of motion 
\be
\partial_i\partial^i |\phi(x)| =
{g^+}\int_{\Sigma}d^2\sigma\delta^6(x-X(\sigma)).\label{EOM-T} 
\ee 
The equations of motion for the self-dual field strength are 
\bea
*d*H^+=J\cr 
*dH^+=J 
\eea 
where the current $J$ is given by 
\bea
J^{\mu\nu}&=&\sum_i g^+\int_{\Sigma_i}dX^{\mu}\wedge dX^{\nu}\delta^6(x-X(\sigma))\cr
&=&\sum_i
g^+\int_{\Sigma_i}d^2\sigma\sqrt{-h}\frac{1}{2!}\varepsilon^{\alpha\beta}\partial_{\alpha}
X^{\mu}\partial_{\beta} X^{\nu}\delta^6(x-X(\sigma)).  
\eea 
We now see that $g^+$ is the electric and the magnetic charge of this string,
which means that we have a self-dual string. The $B$-field from an
infinite string at $X^{1,2,3,4}=0$ thus satisfies the equation of
motion 
\be 
\partial_i\partial^i B^+_{05}(x) =
g^+\int_{\Sigma}d^2\sigma\delta^6(x-X(\sigma)).\label{EOM-B}
\ee 
Both of the equations of motion (\ref{EOM-T}) and (\ref{EOM-B})
reduce to a four-dimensional equation of the form \be
\partial_i\partial^i f({\bf{x}}) = \delta^4({\bf{x}}) \ee which has
the solution \be f({\bf{x}})=f(\infty)-\frac{1}{2\pi^2|{\bf{x}}|^2}.
\ee

These formulas show that the tension decreases as we approach the
position of a string. This is intuitively reasonable if one thinks of
the tension of the string as coming from a membrane stretching between
two five-branes. The string tension will then be proportional to the
distance between the five-branes. We can then understand the
decreasing string tension towards the string as coming from the fact
that a membrane pulls the five-branes closer to each other in the
vicinity of the membrane. The formulas above will break down when we
come close to the string. The formula then says that the tension tends
to minus infinity, which can not be true! The tension must of course
alway be positive. To remedy this one must presumably consider the
exact (still unknown) non-linear interacting quantum theory on the
five-brane. The string action we wrote above in e.g. equation
(\ref{interaction}) is thus only to be seen as an effective action
that is valid only at low energies compared to the masses of the
strings.

The equation of motion for a second string with the same charges as
the first one, which moves in given fields of the tension $T(x)$ and
the potential $B^+_{\mu\nu}(x)$, is most easily derived by varying an
(effective) string action coupled to $B^+$. This action should be
applicable at least at low energies, or when the strings are heavy and
far apart. $B^+$ must of course be on-shell so this action is not useful 
to obtain equations of motion for $B^+$. The strings couple to both $B^+$ 
and its dual potential, but since the field strength is self-dual, the 
dual potential is also $B^+$. We thus get the total coupling as twize a 
pure electrical coupling,
\bea 
&&- \frac{1}{2}\int d^2\sigma T(X)\sqrt{-\gamma}\gamma^{\alpha\beta}
\partial_{\alpha} X^{\mu}\partial_{\beta} X^{\nu}G_{\mu\nu}\cr 
&&- 2g^+\int d^2\sigma\frac{1}{2!}\sqrt{-\gamma}\varepsilon^{\alpha\beta}
\partial_{\alpha} X^{\mu}\partial_{\beta} X^{\nu} B^+_{\mu\nu}(X) 
\eea 
Varying $X^{\mu}$, and then putting the auxiliary metric
$\gamma_{\alpha\beta}$ equal to the induced metric
$h_{\alpha\beta}$, we get 
\bea
T\partial_{\alpha}(\sqrt{-h}h^{\alpha\beta}\partial_{\beta}X_{\rho})
&=&\frac{1}{2}
\sqrt{-h}h^{\alpha\beta}\partial_{\alpha}X^{\mu}\partial_{\beta}X_{\mu}\partial_{\rho}T
-\sqrt{-h}h^{\alpha\beta}\partial_{\alpha}X^{\mu}\partial_{\beta}X_{\rho}\partial_{\mu}T\cr
&& - 2 g^+ \frac{1}{2}\epsilon^{\alpha\beta} H^+_{\mu\nu\rho} \partial_{\alpha}X^{\mu}\partial_{\beta}X^{\nu}
\eea 
which, in the case of a straight string parallel with the first
one, reduces to 
\be
T\partial^{\alpha}\partial_{\alpha}X_{\rho} = \partial_{\rho}T - 2 g^+ \partial_{\rho} B^+_{05} = \frac{2{g^+}^2}{\pi}\frac{x_{\rho}}{|{\bf{x}}|^4}-\frac{2{g^+}^2}{\pi}\frac{x_{\rho}}{|{\bf{x}}|^4}=0.
\ee 
That is, the attractive force due to interaction via scalars
$\phi_a$ cancels the repulsive force via gauge bosons $B^+_{\mu\nu}$
if the two strings are parallel. 

We can also derive this zero force condition from the Hamiltonian. 
For simplicity we assume 
that the only non-zero component of the central charge is $Z_{a}{}^5$ and 
that this component is positive. From the 
anticommutator of two supercharges we can extract the mass as the Hamiltonian 
in the rest frame of the string. We rewrite the mass as follows, where we
introduce a vector field $e$ of unit length:
\bea
M & = & \int_{{\bf{R}}^5}(H^+\wedge *_5 H^+ + d\phi\wedge *_5 d\phi) \cr
  & = & \int_{{\bf{R}}^5}|*_5 H^+ - d\phi_a\wedge e|^2 + \int_{{\bf{R}}^5} 
d^5 x(e^{\mu}\partial_{\mu}\phi)^2 + 2 \int_{{\bf{R}}^5} H^+ \wedge d\phi_a 
\wedge e\cr
  & \geq & 2 |\int_{{\bf{R}}^5} H^+ \wedge d\phi_a \wedge e|
\eea
Here we have equality only if $e^{\mu}\partial_{\mu}\phi = 0$ and 
$*_5 H - d\phi_a\wedge e = 0$. We now let $e$ be a tangent vector field of a 
string. We then define $e$ at an arbitrary point in ${\bf{R}}^5$
 as the vector obtained by parallel transporting $e$ from the string along a 
straigth line ${\bf{R}}^+$ ending on the string. We take the these straight lines to 
be parallel. When we have equality above, $\phi$ is constant along the 
string. We then get
\bea
M & = & 2 |\int_{S^3}H^+ \int_{string}\phi_a e|\cr
  & = & 2 |g^+ \phi_a \int_{string} e|\cr
  & \geq & 2 |g^+ \phi_a \int_{string} dx^5|.
\eea
We thus have BPS saturation, $M = |Z_a{}^5|$, only if $e=dx^5$ (i.e. the 
string is a straight line parallel to the $x^5$-axis), $\phi$ is 
constant along the string and the Bogomolnyi equation 
\be
*_5 H - d\phi_a\wedge dx^5 = 0
\ee
is satisfied. We notice that this equation essentially is the zero force 
condition.

If the strings instead had been
anti-parallel the forces would have added up, since the Lorentz force
would change sign. If one compactifies the $x^5$ direction then
orientation reversal of the string becomes tantamount to changing the
sign of its charge.

\section{Holomorphic factorization of the partition function}

In this section we will consider compact topologically non-trivial
six-manifolds of the form $M_6=S^1 \times M_5$ where $M_5$ is some
compact five-manifold. Having assumed this, it is possible to define a
basis of the homology group $H_3(M_6,{\bf{Z}})$ consisting $A$-cycles
$\{a_i\}$ that wind around the circle and $B$-cycles $\{b_j\}$ that do
not wind around the circle. They are dual in the sense that they have
intersection numbers $a_i\cdot a_j=b_i\cdot b_j=0$, $a_i\cdot
b_j=\delta_i^j$. We will let the circle be in the time-direction, so
in particular the $B$-cycles will be spatial. We define a basis
$[E^i_A]$ and $[E^i_B]$ of $H^3(M_6,{\bf{Z}})$, which is dual to
$H_3(M_6,{\bf{Z}})$, that is,
$\int_{a_i}{E_A}^j=\int_{b_i}{E_B}^j=\delta^j_i$ and
$\int_{b_i}{E_A}^j=\int_{a_i}{E_B}^j=0$. It will be symplectic,
$\int_{M_6}{E_B}^i\wedge {E_A}^j=\delta^j_i$. We will take $E_A$ and
$E_B$ to be harmonic representatives.

We want to make use of a complex structure given by the Hodge duality
operator, *, on the intermediate Jacobian
$H^3(M_6,{\bf{R}})/H^3(M_6,{\bf{Z}})$. \cite{HNS} But this is possible
only if $*^2=-1$. This forces us to make a Wick rotation,
$x^0\rightarrow x^0_E=ix^0$, such that $M_6$ becomes an Euclidean
manifold. We define the period matrix $Z=X + i Y$ with the matrices
$X$ and $Y$ having real entries, by declaring \bea E^+ & = & Z
E_A+E_B\cr E^- & = & \bar{Z} E_A+E_B.  \eea to be self-dual and
anti-self-dual respectively. More explicitly this means that\be
E_B=-XE_A+Y*E_A.  \ee For convenience we have defined our basis forms
such that $\int_{b_i}{E_{\pm}}^j=\delta_i^j$.

Since the harmonic three-forms are on-shell, we can expand the
harmonic parts of the field strengths as \bea H^+_0 & = & {h^+}^t
(ZE_A+E_B)\cr H^-_0 & = & {h^-}^t (\bar{Z}E_A+E_B).  \eea These are
the most general classical solutions on a compact six-manifold. The
quantum oscillations need a different treatment, and will not be
studied in this paper.

The Lagrangian density for a free non-chiral two-form potential
$B_{\mu \nu}$ with Euclidean field strength $H=d B$ is given by \be
{\cal{L}}=-\frac{1}{2\lambda^2}H\wedge *H \ee where $\lambda$ is a
dimensionless coupling constant. When decomposing the field strength
as $H=H^+ + H^-$ where $*H^{\pm}=\pm i H^{\pm}$, we get \be
{\cal{L}}=-\frac{1}{\lambda^2}i H^{-}\wedge
H^{+}=-\frac{iG}{12\lambda^2}\epsilon^{0ijklm}(H^{-}_{0ij}H^{+}_{klm}-H^{-}_{klm}H^{+}_{0ij})
\ee The momentum conjugate to $B_{ij}$ is then (if we temporarily
treat $B_{ij}$ and $B_{ji}$ as independent variables, as in the
appendix) \be
\Pi^{ij}=i\frac{1}{2\lambda^2}\sqrt{G}H^{0ij}=i\frac{i}{12\lambda^2}G\epsilon^{0ijklm}(H^{+}_{klm}-H^{-}_{klm})
\ee If we make the gauge choice $B_{0i}=0$, then the Hamiltonian
density is  \bea {\cal{H}} & = &
i\Pi^{ij}H_{0ij}-{\cal{L}}=-\frac{i}{12\lambda^2}G\epsilon^{0ijklm}(H^{+}_{klm}H^{+}_{0ij}-H^{-}_{klm}H^{-}_{0ij})\cr
& = & -i\frac{1}{\lambda^2}((H^+)_B\wedge (H^+)_A-(H^-)_B\wedge
(H^-)_A) \eea

When we quantize we substitute the Poisson-bracket with a
commutator. We then get the commutation relations \be
[{h^+}^i,{h^+}^j]=0 \ee as we will see in section $4$. We will only be
interested in the zero-mode part of the Hilbert space, which is
spanned by eigenvectors $|h^+>$ of $h^+$. We divide the Hamiltonian
density into a zero-mode part ${\cal{H}}_0$ and a oscillator part
${\cal{H}}_{osc}$. The zero-mode part is given by the operator \be
{\cal{H}}_0=-\frac{i}{\lambda^2}({h^+}^t E_B \wedge E_A^t Z h^+ -
{h^-}^t E_B \wedge E_A^t \bar{Z} h^{-}).  \ee By using the symplectic
property of the three-form basis we get \be \int_{S^1\times M_5}
{\cal{H}}_0=-\frac{i}{\lambda^2}({h^+}^t Z h^+ - {h^-}^t \bar{Z} h^-).
\ee We notice that with Re $Z=0$ this quantity always is positive,
i.e. the energy is positive, as a consequence of the fact that the
period matrix always has the property that Im $Z>0$.

We can extract the zero-mode part by integrating over a spatial cycle
$b^i$, \be \int_{b_i} \frac{H^{\pm}}{g}=\frac{{h^{\pm}}^i}{g}\equiv
{w^{\pm}}^i.  \ee We will define $g$ such that the eigenvalues of \be
w \equiv w^+ + w^- \ee are integers. The minimal magnetic charge is
thus assumed to be $g$. The numerical value of this charge can be
determined from the quantization condition \cite{D} for dyonic strings
in six dimensions with electric and magnetic charges $(e^i,g^i)$, \be
e^ig^j+e^jg^i=2\pi \hbar n^{ij}, \ee where $n^{ij}\in {\bf{Z}}$. This
is a much stronger condition than the corresponding
Dirac-Schwinger-Zwanziger condition in four dimensions, due to the
plus sign. In four dimensions there is a minus sign instead, and hence
one can draw no conclusions by considering two equally charged dyons
in four dimensions. This situation is different in six dimensions. In
particular we can have no theta angle in six dimensions. Another
restriction comes from the fact that any consistent chiral theory must
contain only self-dual strings. By taking two equally charged
self-dual strings with charges $(e^i,g^i)=(g,g)$, the charge
quantization condition implies that the smallest such charge is given
by \be g^2=\pi \hbar.  \ee At this stage it is not clear that
$\int_{b_i} H$, where $H$ is {\sl{non}}-self-dual, should be quantized
in units of $g$. But we will give an argument for this at the end of
this section.

We will now make use of the gauge equivalence of the non-chiral
potential $B \simeq B+\Delta B$, where $\Delta B$ has periods which
are integer multiples of $g$. This fact is derived in appendix $C$ by
using the fact that $B$ is a connection on a gerbe.\footnote{A shorter
argument can be made if the three-cycle is $S^3$. Then we need only
two covers $U_N$ and $U_S$ over each of which the gauge potentials are
uniquely defined, and the complications discussed on triple overlaps
in the appendix do not enter. Let $V_{N(S)}$ be adjacent
neighbourhoods such that $S^3=V_N\cup V_S$. From Stokes theorem we
then get \be \int_{S^3}H=\int_{V_N}dB_N+\int_{V_S}dB_S=\int_{\partial
V_N}(B_N-B_S)=\int_{\partial V_N}\Delta B \ee which indicates that
$\int \Delta B$ over two cycles is quantized in the same units as
$\int H$ is over three-cycles.} The operator which implements such a
gauge transformation is given by  \be \exp{\frac{i}{\hbar}\int_{M^5}
d^5 x \Pi^{ij} \Delta B_{ij}}.  \ee This is proved in the appendix
$B$. Now gauge equivalence means that this operator should have
eigenvalues one. This implies that, if we choose $\Delta B$ such that
it has exactly one non-zero period being $g$ over a two-cycle that has
as its Poincare dual the three-cycle $b_i$, then \be
\frac{i}{\hbar}\int_{M^5} d^5 x \Pi^{ij} \Delta B_{ij} =
-\frac{i}{\hbar\lambda^2} \int_{M^5}(H^+ - H^-)\wedge \Delta B =
-\frac{ig^2}{\hbar\lambda^2} (w^+ - w^-)^i \ee is an integer multiple
$n$ of $2 \pi i$.

Now if we choose our coupling constant such that \be
\frac{g^2}{\lambda^2}=\pi\hbar, \ee which, since $g=\sqrt{\pi\hbar}$,
means that \be \lambda=1, \ee then the zero-mode contribution of the
time-integrated Hamiltonian is \be \int_{S^1\times M^5} {\cal{H}} =
-i\pi({w^+}^t Z w^+ - {w^-}^t \bar{Z} w^-) \ee where
$w^{\pm}=\mp{n}+\frac{w}{2}$ which is necessary in order to
holomorphically factorize the partition function into a finite sum of
chiral times anti-chiral partition functions \cite{HNS}. Each of these
chiral partition functions then describe different chiral
theories. This should allow us to interpret any of these
theta-functions as a trace Tr exp${-TH^+}$ where $H^+$ is (the
zero-mode contribution of) the chiral part of the Hamiltonian and $T$
is an Euclidean time interval. We will take this as a part of the
definition of $H^+$.

We have a gauge invariance in the non-chiral theory, which means that
we can insert an operator which performs such a gauge transformation
without changing the non-chiral partition function. But such an
operator will permute the chiral partition functions. We thus have to
consider the effect of inserting the operator  \be
\exp{\frac{i}{\hbar}\int d^5x 2{\Pi^+}^{ij}\Delta {B^+}_{ij}}.  \ee
which transforms the state $|B^+>$ to $|B^+ + \Delta B^+>$ as is
showed in appendix $B$. The zero-mode contribution to the chiral
partition function is then \bea &&\sum_{w^+}<w^+|
e^{\frac{i}{\hbar}\int d^5x 2{\Pi^+}^{ij}\Delta
{B^+}_{ij}}e^{-\int_{M_6}{\cal{H}}^+} |w^+>\cr
&&=\sum_{w^+}<w^+|e^{\frac{i}{\hbar}2\int H^+\wedge \Delta
B^+}e^{-\int_{M_6}{\cal{H}}^+} |w^+>\cr &&=\sum_{w^+} e^{i 2\pi
{w^+}^t \beta}e^{i\pi w^+ Z w^+} \eea  where we have defined  \be
\beta^i  =  \int_{\tilde{b}_i}\frac{\Delta B^+}{\sqrt{\pi\hbar}} \ee
where $\tilde{b}_i$ is Poincare dual to $b_i$. We do not know any
direct way to deduce over which values $w^+$ should run in the sum
(more than that it should be integer and/or half-integer valued since
it is given by $w^+=n+\frac{w}{2}$). But we know \cite{HNS} that the
answer must be a theta-function $\theta\left[^\alpha_\beta\right](Z)$,
and hence we deduce that ${w^+}^i=\int_{b_i}
\frac{H^+}{\sqrt{\pi\hbar}}\in {\bf{Z}}+\alpha^i$. Thus the `physical'
field strength (by `physical' we will mean a field strength which when
integrated over a three-cycle gives a magnetic charge) is not quite a
connection on a gerbe. We then have to rescale the gauge field as  \be
B_{phys}=\frac{1}{2}\sqrt{\frac{\hbar}{\pi}}B_{math} \ee to obtain the
quantization condition of a self-dual connection, \be \int_{b_i}
\frac{H^+_{math}}{2\pi}\in {\bf{Z}}+\alpha^i.  \ee This is thus the
quantization condition one has in a chiral theory which is
characterized by the $\frac{b_3}{2}$-dimensional vectors $\alpha$ and
$\beta$, with entries in $\frac{1}{2}{\bf{Z}}$. We notice that for
$\alpha=0$, $\int_{b_i} H^+$ is quantized in integer units of the
smallest charge of a self-dual string, $g=\sqrt{\pi\hbar}$. One should
have the same quantization of $\int_{S^3}H^+$ for an $S^3$ around an
infinite magnetically charged string in $\bf{R}^5$ as for an $S^3$ in
a compact manfold that one obtains by a deformation retract of
${\bf{R}}^5-{\bf{R}}={\bf{R}}\times ({\bf{R}}^4-\{0\})={\bf{R}}\times
{\bf{R}}_+ \times S^3$, and by adding some points at infinity to make
the manifold compact.

\section{Commutation relations of surface observables}

In a curved space with Minkowski signature we have the commutation
relations (which are derived in appendix $B$) \cite{H}, \be
[H^+_{ijk}(x),H^+_{i'j'k'}(x')]=i\hbar\frac{3}{2}\epsilon_{i'j'k'[ij}
\partial_{k]} \delta^5(x-x') \ee for the `physical' fields. From this
we can compute the commutation relation between Wilson surfaces
$\int_{D}\frac{H^+}{\sqrt{\pi\hbar}}$ where $D$ is a three-dimensional
surface with boundary $\partial D = \Sigma$,   \bea
&&\left[\int_{D}\frac{H^+}{\sqrt{\pi\hbar}},\int_{D'}\frac{H^+}{\sqrt{\pi\hbar}}\right]\cr
&=&\left[\int_{D}\frac{H^+_{ijk}(x)}{3!\sqrt{\pi\hbar}}dx^i\wedge
dx^j\wedge dx^k,
\int_{D'}\frac{H^+_{i'j'k'}(x')}{3!\sqrt{\pi\hbar}}dx'^i\wedge
dx'^j\wedge dx'^k \right] \cr
&=&-\frac{i}{2\pi}\int_{D'}\frac{1}{3!}dx'^i\wedge dx'^j\wedge dx'^k
\int_{\partial D}\frac{1}{2!}dx^i\wedge dx^j
\epsilon_{i'j'k'ij}\delta^5(x-x')\cr &=&-\frac{i}{2\pi} D' \cdot
\partial D=-\frac{i}{2\pi}L(\Sigma,\Sigma').  \eea The dot, $\cdot$,
denotes the intersection number and $L(\Sigma,\Sigma')$ is defined as
in the last line and is the linking number of the two two-cycles
$\Sigma$ and $\Sigma'$.

We now see that the quantities
${w^+}^i=\int_{b^i}\frac{H^+}{\sqrt{\pi\hbar}}$ commute if $b^i$ are
three-cycles, $\partial b^i=\emptyset$, which justifies our treatment
of these quantities as c-number valued `charges'.

We now consider open curves $D$ with boundary $\Sigma$. Associated
with such surfaces we define the Wilson surface observables  \be
W(\Sigma)\equiv \exp{2\pi i \int_D \frac{H^+}{\sqrt{\pi\hbar}}}.  \ee
By using the BCH-formula we see that these observables commute at
equal time. We could also have gone backwards and showed that the
coupling constant would have to take the value $\lambda=1$ in order
for these surface observables to commute. They should really commute
in order to yield correct commution relations when reducing on a
two-torus. We then get $U(1)$ gauge theory, and these surface
observables become Wilson lines and 't Hooft lines depending on
whether the surface wraps the a- or b-cycle of the two-torus. Then the
above commutation relation reduces to the old fact that the Wilson and
't Hooft lines commute in $U(1)$-gauge theory \cite{tHooft}. We think
it is remarkable that these two entirely different ways of computing
the coupling constant yield the same answer. Using Wilson lines to
compute $\lambda$ did not require a non-trivial topology as
holomorphic factorization did.
 
\section{Reduction to four dimensions}

We will now start from a Minkowski six-manifold and make dimensional
reduction by letting $x^{4,5}\in [0,1]$ be coordinates on a two-torus
and $x^i$ ($i=0,1,2,3$) be the remaining coordinates
\cite{Verlinde}. We will denote the moduli parameter of the torus as
$\tau=\tau_1+i\tau_2$. In this section we will use the mathematicians
conventions for the gauge fields so that they will be connections on a
1-gerbe and 0-gerbe (line-bundle) respectively. This convention has
the advantage that it makes the S-duality transformations look
nicer. We dimensionally reduce the on-shell self-dual field strength
as \bea H^+ & = & \frac{1}{3!}H^+_{ijk}dx^i\wedge dx^j\wedge dx^k \cr
&&+ \frac{1}{2!}F_{ij} dx^i\wedge dx^j\wedge dx^4 +
\frac{1}{2!}\tilde{F}_{ij} dx^i\wedge dx^j\wedge dx^5 \cr &&+
\partial_i B^+_{45} dx^i\wedge dx^4\wedge dx^5.\label{F} \eea Due to
self-duality, $H^+_{ijk}$ and $\partial_i B^+_{45}$ are
related. Likewise $F=dA$ and $\tilde{F}$ are related as \be
\tilde{F}=-\tau_1 F+\tau_2*F \ee if we define $\tau$ as \be
dx^4=\tau_1 dx^5 - \tau_2 * dx^5.  \ee Invariance under
diffeomorphisms implies in particular invariance under modular
transformations of the $T^2$. For $H^+$ to be invariant we must then
impose the following transformation rules, $x^4\rightarrow x^5$,
$x^5\rightarrow -x^4$, $\tau\rightarrow -\frac{1}{\tau}$,
$F\rightarrow \tilde{F}$, $\tilde{F}\rightarrow -F$ and
$x^4\rightarrow x^4+x^5$, $\tau\rightarrow \tau+1$,
$\tilde{F}\rightarrow \tilde{F}-F$. This diffeomorphism invariance is
S-duality from the four-dimensional point of view.
 
We will now integrate $H^+$ over three-cycles $\Sigma\times\gamma$
where $\Sigma$ is a two-cycle not on the torus, and $\gamma$ is either
the $a$ or the $b$-cycle of the torus, normalized such that $\int_a
dx^4=\int_b dx^5=1$, $\int_a dx^5=\int_b dx^4=0$. We then get \be
\left(
\begin{array}{c}
\int_{\Sigma\times a}H^+ \\ \int_{\Sigma\times b}H^+
\end{array}
\right) = \left(
\begin{array}{c}
\int_{\Sigma}F \\ \int_{\Sigma}\tilde{F}
\end{array}
\right) = \left(
\begin{array}{cc}
1 & 0 \\ -\tau_1 & \tau_2
\end{array}
\right) \left(
\begin{array}{c}
\int_{\Sigma}F \\ \int_{\Sigma}*F
\end{array}
\right).  \ee In section 3 we saw that $\int_{\Sigma\times
\gamma}\frac{H^+}{2\pi}= w_{\gamma}$ where $w^+_{\gamma}$ is either in
$\bf{Z}$ or in ${\bf{Z}}+\frac{1}{2}$ depending on which theory we are
looking at (i.e. on which theta function we pick). Now we should
choose the theory which corresponds to the theta function with zero
characteristics, $\theta\left[^{00\cdots 0}_{00\cdots 0}\right]$,
which was found in \cite{AG} to be the only theory which candidate to
be modular invariant on manifolds of the form $\tilde{T^2}\times M_4$
provided that we choose our $A$- and $B$-cycles properly. Here we
should consider the case when $M_4=S^1\times M_3$ where $S^1$ is
(Euclidean and periodic) time. We then combine one of the one-cycles
of $T^2$ with the $S^1$-time to a new two-torus $\tilde{T^2}$. The
remaining four-manifold will then contain a one-cycle. This means that
the modular group of the $\tilde{T^2}$ does not constrain all the
entries in $\alpha$ and $\beta$ to be zero. But by combining modular
groups from all two-tori with one cycle being the $S^1$-time (in the
case that $M_4=S^1\times M_3$ with $M_3$ simply connected, we have the
two two-tori $\tilde{T^2}=S^1\times a$ and
$\tilde{\tilde{T^2}}=S^1\times b$) we find that all entries in
$\alpha$ and $\beta$ must be zero.

In our mathematical convention the four-dimensional magnetic and
electric charges will, as we will see below, be given as \bea
g=\sqrt{\frac{\tau_2\hbar}{2\pi}}\int_{\Sigma} F\\
q=\sqrt{\frac{\tau_2\hbar}{2\pi}}\int_{\Sigma} *F.  \eea Under
$\tau\rightarrow\tau+1$ we want the charges of a dyon to transform as
$(g,q)\rightarrow (g,q+e)$ where $e$ is the smallest electric charge
unit. In the case when $\tau_1=0$, we want $(g,q)\rightarrow (q,-g)$
under $\tau_2\rightarrow \frac{1}{\tau_2}$. This explains why we have
to insert a factor proportional to $\sqrt{\tau_2}$. Now we find that
\be \left(
\begin{array}{c}
g \\ q
\end{array}
\right) = \sqrt{\frac{2\pi\hbar}{\tau_2}} \left(
\begin{array}{cc}
\tau_2 & 0 \\ \tau_1 & 1
\end{array}
\right) \left(
\begin{array}{c}
w^+_a \\ w^+_b
\end{array}
\right) \ee If we now put $\tau=\frac{\theta}{\pi}+\frac{8\pi
i}{g_{YM}^2 \hbar}$, then we get \bea g & = &
\frac{4\pi}{g_{YM}}w^+_a\cr q & = & \frac{g_{YM}\hbar}{2}w^+_b +
\frac{\theta g_{YM}\hbar}{2\pi}w^+_a \eea The smallest electric charge
is then $e=\sqrt{\frac{2\pi\hbar}{\tau_2}}=\frac{g_{YM}\hbar}{2}$.

This is a charge quantization that one has on a topologically
non-trivial four-manifold in abelian $N=4$ SYM with coupling constant
$g_{YM}$ and theta angle $\theta$. One also has this charge
quantization on a topologically trivial four-manifold (apart from
monopole singularities) and with gauge group $SU(2)$ (or the dual
group $SO(3)$) spontaneously broken to $U(1)$ by a Higgs vacuum
expectation value \cite{Witten79}. What we have obtained above is the
abelian SYM since we did not consider any strings in six
dimensions. To obtain $SU(2)$ (spontaneously broken to $U(1)$) SYM one
should consider a six dimensional theory with one tensor multiplet
coupled to massive strings which wind around non-trivial one-cycles in
the six-manifold \cite{GLPT}. The same calculation as we did above
goes through also in this case, only the interpretations differ. The
six-manifold is then a four-manifold times a two-torus with a
self-dual string winding around the $a$- and $b$-cycles. We notice
that the three-cycle $\Sigma\times a$ encloses a $b$-cycle that sits
at some point within $\Sigma$. So $\int_{\Sigma\times a}H^+$ is now
the magnetic charge of a string that winds around the $b$-cycle and is
located at this particular point within $\Sigma$. This setup, with one
tensor-multiplet and tensile strings winding around two-tori, reduces
to a spontaneosly broken $SU(2)$ gauge theory in four dimensions - the
massive string reduces to the massive gauge fields in four dimensions
(which have obtained their mass from a Higgs mechanism which has
broken $SU(2)$ down to $U(1)$) \cite{HG}. More generally a theory with
$N-1$ massless tensor multiplets should reduce to a theory with $N-1$
massless $U(1)$ gauge fields, each one being associated with a Cartan
generator of the gauge group, so we then obtain an $SU(N)$ gauge
theory spontaneously broken by a generic Higgs field to $U(1)^{N-1}$.

This also has an interpretation in terms of $N$ parallel M5-branes,
with M2-branes stretching between them. When all the M5-branes are
well separated from each other (which means that we consider an
effective theory at low energies, compared to the masses of the
strings) we have, on each M5-brane, a theory with one massless tensor
multiplet coupled to tensile strings. So when considering all the
M5-branes we should have a theory with N $U(1)$ gauge fields, or if
one so wish, one gauge field transforming under $U(1)^N$. One of these
$U(1)$ factors describes the center of mass motion of the whole system
of branes, so the internal theory is a $U(1)^{N-1}$ gauge theory
coupled to tensile strings. When the M5-branes come close to each
other (or when we consider the theory at high energies that become
comparable with the mass of the strings) no-body knows how to describe
the six-dimensional theory, but when one compactifies it to four
dimensions by e.g. letting the strings wind around a two-torus we
expect that one gets $SU(N)$ SYM in four dimensions.

\vskip 0.5truecm I would like to thank M. Henningson for discussions.

\newpage
\appendix
\section{Appendix - Gamma matrices}

\subsection{The SO(1,5) spinor representaion}

The chirality matrix is $\gamma=-\gamma_0\gamma_1\cdots \gamma_5$. We
let $c^{AB}$ denote the charge conjugation matrix. It can be choosen
to be either symmetric or antisymmetric. We will choose it to be
symmetric. From the fact that $4\times 4'$ contains a singlet we
deduce that the invariant charge conjugation tensor must be of the
form \be c^{AB}=\left(
\begin{array}{cc}
0 & c^{\alpha\beta'} \\ c^{\alpha'\beta} & 0
\end{array}
\right) \ee where $c^{\alpha\beta'}=c^{\beta'\alpha}$, if we choose a
representation where  \bea \gamma_A{}^B=\left(
\begin{array}{cc}
\delta_{\alpha}^{\beta} & 0 \\ 0 & -\delta_{\alpha'}^{\beta'}
\end{array}
\right)\cr (\gamma_{\mu})_A{}^B=\left(
\begin{array}{cc}
0 & (\gamma_{\mu})_{\alpha}{}^{\beta'} \\
(\gamma_{\mu})_{\alpha}{}^{\beta'} & 0
\end{array}
\right) \eea The gamma matrices must be antisymmetric, e.g
$(\gamma^{\mu})_{\alpha\beta}=-(\gamma^{\mu})_{\beta\alpha}$. We will
raise and lower spinor indices as  \bea
\psi^{\alpha}=c^{\alpha\beta'}\psi_{\beta'}\cr
\psi^{\alpha}=\psi^{\beta'}c_{\beta'\alpha} \eea where \be
c_{\alpha\beta'}c^{\beta'\gamma}=\delta_{\alpha}^{\gamma}.  \ee

\subsection{The SO(5) spinor representation}

We let $\Omega^{ij}=-\Omega^{ji}$ denote the charge conjugation
matrix, and we use the conventions \bea \psi^i &\equiv&
\Omega^{ij}\psi_j\cr \psi_i &\equiv& \psi^j\Omega_{ji}\cr \psi_i\chi^i
&=& \psi^i\Omega_{ij}\chi^j = -\psi^i\chi_i \eea Then $\Omega$ has to
satisfy \be \Omega^{ij}\Omega_{jk}=\Omega_{kj}\Omega^{ji}=-\delta^i_k.
\ee We will denote the gamma matrices as $(\sigma_a)_i{}^j$.

\section{Appendix - Canonical quantization}

We will here quantize the non-chiral theory with Lagrangian density
${\cal{L}}=-\frac{1}{2}H\wedge*H$, $H=dB$, in a curved space with
Minkowski signature, and then we will also quantize the corresponding
chiral theory. We will first treat $B_{ij}$ and $B_{ji}$ as
independent fields. It is only the antisymmetric part,
$\frac{1}{2}(B_{ij}-B_{ji})$, which occurs in the action. We then get
the primary constraints for the canonical momentum
$\Pi^{ij}=-\frac{\sqrt{|G|}}{2}H^{0ij}$ associated with $B_{ij}$,
$\Pi^{ij}+\Pi^{ji}=0$ and $\Pi^{i0}=0$, and the secondary constraints
$\partial_i \Pi^{ij}=0$. We eliminate the symmetric parts by the gauge
fixing conditions $B_{ij}+B_{ji}=0$, and the $0i$-components by the
gauge fixing condition $B_{i0}=0$. By imposing the gauge fixing
condition $\partial^i B_{ij}=0$ we have finally fixed the gauge
completely. The Poisson bracket is as always given by \be
\{B_{ij}(x),\Pi^{i'j'}(x')\}=\delta_i^{i'}\delta_j^{j'} \ee and for
the (partially) reduced phase space variables we get the bracket
(which rigorously should be computed as a Dirac-bracket. The result
happens to coincides with what one gets by just antisymmetrizing the
indices), \be \{B_{[ij]}(x),\Pi^{[i'j']}(x')\}_*=\delta_{ij}^{i'j'}.
\ee Now, after that we have reduced our phase space, we will drop the
antisymmetrization symbol $[$  $]$.

The constraints we have choosen here are not independent. There are
two relations between them, $\partial_i\partial_j\Pi^{ij}=0$ and
$\partial^i\partial^j B_{ij}=0$. We therefore introduce two $4\times
5$-matrices $\alpha$ and $\beta$ of rank $4$. Then the independent
second class constraints can be expressed as \bea \alpha_k^I
\partial_i \Pi^{ik}&=&0\cr \beta^{k'}_{I'} \partial^{i'} B_{i'k'}&=&0
\eea where $I,I'=1,2,3,4$. The matrices $\alpha$ and $\beta$ can not
be any rank $4$ matrices. They are constrained by the condition that
none of the above constraints are trivially fulfilled. In flat space
we can work in the Fourier space. There we see that $\alpha_k^K(k_i)$
must be orthogonal to the vector space spanned by the momentum vector
$k_i$ (and similarly for $\beta$). Now the dimension of this space
coincides precisely with the rank of $\alpha$, so that such a matrix
$\alpha$ (and similarly $\beta$) exists.

When we quantize we shall substitite the Dirac bracket, $\{$ $,$
$\}_{**}$, on the fully reduced phase space, by the anticommutator
$\frac{1}{i\hbar}[$ $,$ $]$. We thus have to compute the Dirac
bracket, which is given by \bea &&\{B_{ij}(x),\Pi^{i'j'}(x')\}_{**} =
\{B_{ij}(x),\Pi^{i'j'}(x')\}_*\cr && - \int d^5 y d^5 y'
\{B_{ij}(x),\alpha_l^L\partial_k \Pi^{kl}(y)\}_*\cr
&&C^{-1}(y,y')_{L}{}^{L'}\{\beta^{l'}_{L'}\partial^{k'}B_{k'l'}(y'),\Pi^{i'j'}(x')\}_*
\eea Here $C(y,y')^L{}_{L'}=\alpha_l^L\beta^{l'}_{L'}\{\partial_k
\Pi^{kl}(y),\partial^{k'}B_{k'l'}(y')\}$, the exact form of which will
be of no use for our purposes. By integrating by parts we get \be
\{B_{ij}(x),\Pi^{i'j'}(x')\}_{**} = \{B_{ij}(x),\Pi^{i'j'}(x')\}_* -
\delta_{ij}^{kl}\delta_{k'l'}^{i'j'}\partial_k
\partial^{k'}D(x,x')_l{}^{l'} \ee for some continuous functions
$D(x,x')_l{}^{l'}$. Canonical quantization means that we should put
\be
[B_{ij}(x),\Pi^{i'j'}(x')]=i\hbar\left[\delta_{ij}^{i'j'}-\delta_{ij}^{kl}\delta_{k'l'}^{i'j'}\partial_k
\partial^{k'}D(x,x')_l{}^{l'}\right].  \ee and hence \bea
&&[H_{ijk}(x),H^{0lm}(y)]=-\frac{6}{\sqrt{|G(y)|}}\partial_{[i}[B_{jk]}(x),\Pi^{lm}(y)]\cr
&&=-i\hbar\frac{6}{\sqrt{|G(y)|}}\delta^{lm}_{[jk}\partial_{i]}\delta^5(x-y)
\eea where in the last step we have noticed that
$\partial_{[i}\partial_{j]}=0$. This implies that \be
[{H^+}_{ijk}(x),{H^+}^{0lm}(y)]=-i\hbar\frac{3}{\sqrt{|G|}}\delta_{[jk}^{lm}\partial_{i]}\delta^5
(\bf{x-y}) \ee or equivalently, \be
[H^+_{ijk}(x),H^+_{lmn}(y)]=i\hbar\frac{3}{2}\varepsilon_{[lmnij}\partial_{k]}\delta^5
(\bf{x-y}).  \ee In the case that the fields are on-shell we can go
one step further and rewrite this as \be
=\frac{6}{\sqrt{|G(y)|}}\partial_{[i}[B^+_{jk]}(x),{\Pi^-}^{lm}(y)]
\ee where we have divided the conjugate momentum into a chiral and a
anti-chiral part as \be {\Pi^{\pm}}^{ij}\equiv
-\frac{1}{2}\sqrt{|G|}{H^{\pm}}^{0ij} \ee From this we deduce that \be
[B^{\pm}_{ij}(x),{\Pi^{\pm}}^{i'j'}(x')]=i\hbar\left[\frac{1}{2}\delta_{ij}^{i'j'}+\delta_{ij}^{kl}\delta_{k'l'}^{i'j'}\partial_k
\partial^{k'}\tilde{D}(x,x')_l{}^{l'}\right], \ee where
$\tilde{D}_l{}^{l'}$ are some continuous functions.

We finally show that $\exp{\frac{i}{\hbar}\int \Pi^{ij}\Delta B_{ij}}$
translates $B$ to $B+\Delta B$ provided that $\Delta B_{ij}$ obey the
gauge fixing constraints $\partial^{k}\Delta B_{kl}=0$, \bea
&&\frac{i}{\hbar}[\int d^5x' \Pi^{i'j'}(x')\Delta
B_{i'j'}(x'),B_{ij}(x)]\cr &&=\Delta B_{ij}(x)-\int d^5x'
\delta_{ij}^{kl}[\partial_k \partial^{k'} D(x,x')]\Delta
B_{k'l'}(x')\cr &&=\Delta B_{ij}(x)+\int d^5x
\delta_{ij}^{kl}[\partial_k D(x,x')]\partial^{k'}\Delta
B_{k'l'}(x')\cr &&=\Delta B_{ij}(x).  \eea

\section{Appendix - The Dirac quantization condition and the Wilson surface}

We will here obtain the Dirac quantization condition on manifolds with
arbitrary topology. This we do by straightforwardly generalizing the
arguments in \cite{AG}. We then let $b$ be any three-cycle which we
cover by contractible neighbourhoods $U_{\alpha}$ with no more than
quadruple overlaps. We will assume the overlap regions to be
contractible and let $V_{\alpha}$ be adjacent neighbourhoods obtained
by contracting the overlaps. We will indicate orientation reversal
with minus signs. We will define the common boundary surface $\partial
V_{\alpha}\cap \partial V_{\beta}=V_{\alpha}\cap V_{\beta}$ (we could
remove $\partial$ since these neighbourhoods were adjacent) of the
boundaries $\partial V_{\alpha}$ and $\partial V_{\beta}$ to be
antisymmetric in $\alpha$ and $\beta$. We denote the intersection line
between two such common boundaries by  \be
l_{\alpha\beta\gamma}=-\partial(\partial V_{\alpha}\cap \partial
V_{\beta}) \cap\partial(\partial V_{\alpha}\cap \partial
V_{\gamma})=V_{\alpha}\cap V_{\beta}\cap V_{\gamma} \ee It is totally
antisymmetric. If $U_{\alpha}$ has a overlap only with
$U_{\beta},U_{\gamma}$ and $U_{\delta}$, then
$\emptyset=\partial\partial V_{\alpha}=\partial(\partial V_{\alpha}
\cap \partial V_{\beta})+\partial(\partial V_{\alpha}\cap \partial
V_{\gamma})+\partial(\partial V_{\alpha}\cap \partial V_{\delta})$ and
we find that  \bea \partial(\partial V_{\alpha}\cap \partial
V_{\beta})=l_{\alpha\beta\gamma}+l_{\alpha\beta\delta}.  \eea
Similarly we can compute that \be \partial
l_{\alpha\beta\gamma}=-\partial l_{\alpha\beta\delta}=\partial
l_{\alpha\gamma\delta}=-\partial l_{\beta\gamma\delta}=V_{\alpha}\cap
V_{\beta}\cap V_{\gamma}\cap V_{\delta} \ee which is a finite set of
points.

We note that $dB_{\alpha}=dB_{\beta}$ in $U_{\alpha}\cap
U_{\beta}$. We can therefore use the Poincare lemma to obtain \be
B_{\alpha}-B_{\beta}=dA_{\alpha\beta}.  \ee in $U_{\alpha}\cap
U_{\beta}$. Similarly we see that
$d(A_{\alpha\beta}+A_{\beta\gamma}+A_{\gamma\alpha})=0$ in
$U_{\alpha}\cap U_{\beta}\cap U_{\gamma}$ and so, by the Poincare
lemma, we can write \be
A_{\alpha\beta}+A_{\beta\gamma}+A_{\gamma\alpha}=df_{\alpha\beta\gamma}
\ee in $U_{\alpha}\cap U_{\beta}\cap U_{\gamma}$.

Now we have all ingredients to compute the period of the field
strength $H=dB$, \bea &&\int_{b} H
=\sum_{\alpha}\int_{V_\alpha}dB_{\alpha} =\sum_{\alpha}\int_{\partial
V_{\alpha}}B_{\alpha}\cr &&=\sum_{\alpha<\beta}\int_{\partial
V_{\alpha}\cap\partial V_{\beta}}(B_{\alpha}-B_{\beta})
=\sum_{\alpha<\beta}\int_{\partial V_{\alpha}\cap\partial
V_{\beta}}dA_{\alpha\beta} =\sum_{\alpha<\beta}\int_{\partial
(\partial V_{\alpha}\cap\partial V_{\beta})}A_{\alpha\beta}\cr
&&=\sum_{\alpha<\beta<\gamma}\int_{l_{\alpha\beta\gamma}}(A_{\alpha\beta}
+ A_{\beta\gamma} + A_{\gamma\alpha})
=\sum_{\alpha<\beta<\gamma}\int_{l_{\alpha\beta\gamma}}
df_{\alpha\beta\gamma}\cr &&=\sum_{U_{\alpha}\cap U_{\beta}\cap
U_{\gamma}\cap U_{\delta}\cap \Sigma}
(f_{\beta\gamma\delta}-f_{\alpha\gamma\delta}+f_{\alpha\beta\gamma}-f_{\alpha\beta\gamma})
\eea Now, by definition of a connection on a gerbe, if
$\frac{2\pi}{g}B$ is such a connection, then
$f_{\beta\gamma\delta}-f_{\alpha\gamma\delta}+f_{\alpha\beta\gamma}-f_{\alpha\beta\gamma}\in
g{\bf{Z}}$.

Now we turn to the Wilson surface. It should be something like a
two-form $B$ integrated over a two-cycle $\tilde{b}$. If $\tilde{b}$
is covered by two neighbourhoods and $\gamma_{\alpha}$ and
$\gamma_{\beta}$ are adjacent neighbourhoods then we consider the
quantity \be
\int_{\gamma_{\alpha}}B_{\alpha}+\int_{\gamma_{\beta}}B_{\beta}.  \ee
This changes when we deform the neighbourhoods $\gamma_{\alpha}$ and
$\gamma_{\beta}$ such that
$\delta(\gamma_{\alpha}+\gamma_{\beta})=0$. We get the variation \be
\int_{\delta\gamma_{\alpha}}(B_{\alpha}-B_{\beta})=\int_{\delta\gamma_{\alpha}}dA_{\alpha\beta}=\int_{\partial\delta\gamma_{\alpha}}A_{\alpha\beta}
\ee which we also can write as \be
\int_{\delta\partial\gamma_{\alpha}}A_{\alpha\beta} \ee So a sensible
definition of a Wilson surface of a two-cycle which can be covered by
at most two neighbourhoods is  \be
\int_{\gamma}B=\int_{\gamma_{\alpha}}B_{\alpha}+\int_{\gamma_{\beta}}B_{\beta}-\int_{\partial\gamma_{\alpha}\cap
\partial \gamma_{\beta}}A_{\alpha\beta} \ee In order to understand
what happens for a manifold which has to be covered by three
neighbourhoods we make a variation such that
$\delta(\gamma_{\alpha}+\gamma_{\beta}+\gamma_{\gamma})=0$ and compute
the variation \bea &&\delta \left(
\int_{\gamma_{\alpha}}B_{\alpha}+\int_{\gamma_{\beta}}B_{\beta}+\int_{\gamma_{\gamma}}B_{\gamma}\right.\cr
&&\left.-\int_{\partial\gamma_{\alpha}\cap\partial\gamma_{\beta}}A_{\alpha\beta}
-\int_{\partial\gamma_{\alpha}\cap\partial\gamma_{\gamma}}A_{\alpha\gamma}
-\int_{\partial\gamma_{\beta}\cap\partial\gamma_{\gamma}}A_{\beta\gamma}
\right)\cr
&=&\int_{\delta\gamma_{\alpha}}(B_{\alpha}-B_{\gamma})+\int_{\delta
\gamma_{\beta}}(B_{\beta}-B_{\gamma})\cr
&&-\int_{\delta(\gamma_{\alpha}\cap
\gamma_{\beta})}A_{\alpha\beta}-\int_{\delta(\gamma_{\alpha}\cap
\gamma_{\gamma})}A_{\alpha\gamma}-\int_{\delta(\gamma_{\beta}\cap
\gamma_{\gamma})}A_{\beta\gamma}\cr
&=&\int_{\delta\partial\gamma_{\alpha}}A_{\alpha\beta}+\int_{\delta\partial
\gamma_{\beta}}A_{\beta\gamma}\cr &&-\int_{\delta(\gamma_{\alpha}\cap
\gamma_{\beta})}A_{\alpha\beta}-\int_{\delta(\gamma_{\alpha}\cap
\gamma_{\gamma})}A_{\alpha\gamma}-\int_{\delta(\gamma_{\beta}\cap
\gamma_{\gamma})}A_{\beta\gamma}\cr
&=&\int_{\delta(\partial\gamma_{\alpha}\cap\partial\gamma_{\beta})}
(-A_{\alpha\beta}-A_{\beta\gamma}-A_{\gamma\alpha})
=\int_{\delta(\partial\gamma_{\alpha}\cap\partial\gamma_{\beta}\cap\partial\gamma_{\gamma})}-df_{\alpha\beta\gamma}\cr
&=&-\int_{\partial\delta(\partial\gamma_{\alpha}\cap\partial\gamma_{\beta}\cap\partial\gamma_{\gamma})}f_{\alpha\beta\gamma}
=-\delta\int_{\partial\gamma_{\alpha}\cap\partial\gamma_{\beta}\cap\partial\gamma_{\gamma}}f_{\alpha\beta\gamma}
\eea We have then noticed that if
$\partial(\gamma_{\alpha}+\gamma_{\beta}+\gamma_{\gamma})=\emptyset$,
then
$\partial(\gamma_{\alpha}\cap\gamma_{\beta})=\gamma_{\alpha}\cap\gamma_{\beta}\cap\gamma_{\gamma}$
since these neighbourhoods were assumed to be adjacent\footnote{Of
course
${\partial\gamma_{\alpha}\cap\partial\gamma_{\beta}\cap\partial\gamma_{\gamma}}$
is just a set of points. The integral then means that one should
evalute the integrand in those points.}. A sensible definition on
manifolds of arbitrary topology of a Wilson surface is thus \be
\int_{\tilde{b}}B=\sum_{\alpha}\int_{\gamma_{\alpha}}B_{\alpha}-\sum_{\alpha<\beta}\int_{\partial\gamma_{\alpha}\cap\partial\gamma_{\beta}}A_{\alpha\beta}+\sum_{\alpha<\beta<\gamma}\int_{\partial\gamma_{\alpha}\cap\partial\gamma_{\beta}\cap\partial\gamma_{\gamma}}f_{\alpha\beta\gamma}\label{Wilson}
\ee which thus is independent of how we deform the boundaries of our
adjacent neighbourhoods which cover the 2-cycle $\tilde{b}$.

This definition is also nice in that it gives the same value on the
Wilson surface for such two-cycles which can be covered by two
neighbourhoods, as it does if we instead cover it by three
neighbourhoods. But if we add a fourth neighbourhood in such a way
that we get a quadruple overlap, then the Wilson surface
changes. Fortunately it changes in a well-behaved way as we will see
now. The simplest way to compute the change is to make use of the fact
that we can continuously deform our fourth curve piece at our wish,
without changing the value of the Wilson line in the way we have
constructed it. We therefore choose this new curve piece
$\gamma_{\delta}$ in such a way that it shrinks against the point
$\gamma_{\alpha}\cap\gamma_{\beta}\cap\gamma_{\delta}$ so that
$\gamma_{\alpha}\cap\gamma_{\beta}\cap\gamma_{\delta}=\gamma_{\beta}\cap\gamma_{\gamma}\cap\gamma_{\delta}=\gamma_{\gamma}\cap\gamma_{\alpha}\cap\gamma_{\delta}$. We
can then express the change of the Wilson surface as \bea
&&\int_{\alpha\cap\beta\cap\delta}f_{\alpha\beta\delta}
+\int_{\alpha\cap\gamma\cap\delta}f_{\alpha\gamma\delta}
+\int_{\beta\cap\gamma\cap\delta}f_{\beta\gamma\delta}
-\int_{\alpha\cap\beta\cap\delta}f_{\alpha\beta\gamma}\cr
&&=\int_{\alpha\cap\beta\cap\delta}(-f_{\alpha\beta\gamma}+f_{\alpha\beta\delta}-f_{\alpha\gamma\delta}+f_{\beta\gamma\delta})
\eea Now this quantity is an integer multiple of $g$. To obtain a
uniquely defined Wilson surface observable we should then
exponentiate, \be \exp\left\{\frac{2\pi
i}{g}\int_{\tilde{b}}B\right\}.  \ee There is a also different way to
express the Wilson surface \cite{Hitchin}. One notices that the
three-form field strength $H=dB$, when pulled back to the two-cycle
$\tilde{b}$, necessarily is zero. We now cover $\tilde{b}$ with
neighbourhoods $U_{\alpha}$. Since now $dB_{\alpha}=0$ in
$U_{\alpha}$, we can write $B_{\alpha}=d\Lambda_{\alpha}$. Furthermore
$B_{\alpha}-B_{\beta}=dA_{\alpha\beta}$ in $U_{\alpha}\cap U_{\beta}$
so we can write
$A_{\alpha\beta}-(\Lambda_{\alpha}-\Lambda_{\beta})=df_{\alpha\beta}$. Finally
we get, in $U_{\alpha}\cap U_{\beta}\cap U_{\gamma}$, that
$df_{\alpha\beta\gamma}=A_{\alpha\beta}+A_{\beta\gamma}+A_{\gamma\alpha}=d(f_{\alpha\beta}+f_{\beta\gamma}+f_{\gamma\alpha})$,
so $c_{\alpha\beta\gamma}\equiv
f_{\alpha\beta}+f_{\beta\gamma}+f_{\gamma\alpha}-f_{\alpha\beta\gamma}$
is constant. Now if we compute the Wilson surface as we have defined
it in (\ref{Wilson}), and use that $B_{\alpha}=d\Lambda_{\alpha}$ in
$\gamma_{\alpha}$ and
$A_{\alpha\beta}=df_{\alpha\beta}+(\Lambda_{\alpha}-\Lambda_{\beta})$
on $\gamma_{\alpha}\cap\gamma_{\beta}$, we find that
$\int_{\tilde{b}}B=-\sum_{\alpha<\beta<\gamma}\int_{\partial\gamma_{\alpha}\cap\partial\gamma_{\beta}\cap\partial\gamma_{\gamma}}c_{\alpha\beta\gamma}$. Since
the `generator' $\frac{2\pi}{g}f_{\alpha\beta\gamma}$ of the U(1)
gauge transformations is well-defined only modulo $2\pi{\bf{Z}}$ we
immediately see that $\int_{\tilde{b}}B$ is well-defined only modulo
$g{\bf{Z}}$.


\vskip 0.5truecm


\newpage
\begin{thebibliography}{999}
\bibitem{Witten95}E. Witten, `Some Comments on string dynamics',
hep-th/9507121.
\bibitem{OW}E. Witten, D. Olive, `Supersymmetry algebras that include
topological charges', Phys. Lett. B78 (1978) 1
\bibitem{Witten79}E. Witten, `Dyons of charge $e\theta/2\pi$',
Phys. Lett. B86 (1979) 3,4.
\bibitem{HLW}P. S. Howe, N. D. Lambert, P. C. West, `The threebrane
soliton of the M-fivebrane', hep-th/9710033.
\bibitem{GGT}J. P. Gauntlett, J. Gomis, P. K. Townsend, `BPS bounds
for worldvolume branes', hep-th/9711205.
\bibitem{CKP}P. Claus, R. Kallosh, A. Proeyen, `M5-brane and
superconformal (0,2) tensor multiplet in 6 dimensions', hep-th/9711161.
\bibitem{BSP}E. Bergshoeff, E. Sezgin, A. Proeyen, `(2,0) tensor
multiplets and conformal supergravity in D=6.', hep-th/9904085.
\bibitem{HNS}M. Henningson, B. Nilsson, P. Salomonson, `Holomorphic
factorization of correlation functions in (4k+2)-dimensional (2k)-form
gauge theory' hep-th/9908107.
\bibitem{G}A. Gustavsson, `On the holomorphically factorized partition
function for abelian gauge theory in six dimensions', hep-th/0008161.
\bibitem{Bonelli}G. Bonelli, `On the supersymmetric index of the
M-theory 5-brane and little string theory', hep-th/0107051.
\bibitem{H}M. Henningson, `Commutation relations for surface
observables in six-dimensional $(2,0)$ theory', hep-th/0012070.
\bibitem{tHooft}G. 't Hooft, `On the phase transition towards
permanent quark confinement', Nucl. Phys. B138 (1978) 1.
\bibitem{Verlinde}E. Verlinde, `Global aspects of electric-magnetic
duality', hep-th/9506011.
\bibitem{HG}M. Henningson, A. Gustavsson, `A short representation of
the six-dimensional (2,0) algebra', hep-th/0104172.
\bibitem{AG}M. Alvarez, D. Olive `The Dirac quantization for fluxes on
four-manifolds', hep-th/9906093.
\bibitem{D}S. Deser, A. Gomberoff, M. Henneaux, C. Teitelboim,
`p-brane dyons and electric-magnetic duality', hep-th/9712189.
\bibitem{Hitchin}N. Hitchin, `Lectures on special lagrangian
submanifolds', math.DG/9907034.
\bibitem{GLPT}M.B. Green, N.D. Lambert, G. Papadopoulos, P.K. Townsend
`Dyonic p-branes from self-dual (p+1)-branes', hep-th/9605146.
\end{thebibliography}
\end{document}